\documentclass[preprint,showpacs,preprintnumbers,amsmath,amssymb]{revtex4}

\usepackage{graphicx}% Include figure files
\usepackage{dcolumn}% Align table columns on decimal point
\usepackage{bm}% bold math

\begin{document}

\title{Exact coherent states of a noninteracting Fermi gas in a harmonic
trap}

\author{Dae-Yup Song}
 \email{dsong@sunchon.ac.kr}
\affiliation{Department of Physics, Sunchon National University,
Jeonnam 540-742, Korea}

\date{\today}% It is always \today, today,

\begin{abstract}
Exact and closed-form expressions of the particle density, the
kinetic energy density, the probability current density, and the
momentum distribution are derived for a coherent state of a
noninteracting Fermi gas, while such a state can be obtained from
the ground state in a $d$-dimensional isotropic harmonic trap by
modulating the trap frequency and shifting the trap center.
Conservation laws for the relations of the densities are also given.
The profile of the momentum distribution turns out to be identical
in shape with that of the particle density, however,
%as an observable manifestation of the uncertainty principle,
the dispersion of the distribution increases (decreases) when that
of the particle density is decreased (increased). The expressions
are also applicable for a sudden and total opening of the trap, and
it is shown that, after the opening, the gas has a stationary
momentum distribution whose dispersion could be arbitrarily large or
small.
%Conservation laws for the relations of the densities are
%also given.
\end{abstract}

 \maketitle

Experiments on quantum degenerate Fermi atoms provide motivation for
the theoretical analysis of the noninteracting Fermi gas in a
harmonic trap. Since the interaction energy per particle is small
compared with the Fermi energy, the interactions between atoms are
controllable, and since the short-range interaction may be ignored
in the long-term behavior of the expanding Fermi gas, theoretical
studies of noninteracting particles are important for understanding
experiments on degenerate Fermi gases \cite{Mueller}. Indeed, in the
deep BCS limit, the Fermi gases of atoms are considered to be
noninteracting \cite{RGGHJ,Stringari}. As most experiments are
performed in harmonic traps, analytic expressions of the particle
density and the kinetic energy density have been given for the
$(M+1)-$shell filled ground state of the Fermi gas in a
$d$-dimensional isotropic harmonic trap \cite{BZ}.

For an isotropic harmonic oscillator, the time-dependent
Schr\"{o}dinger equation is invariant under the displacement-type
transformation (DTT), and under squeeze-type transformation (STT)
with a rescaling of time. As implied by Kohn's theorem, the system
of a single-component gas is still invariant under the extended DTT,
which can be used to find a coherent state of the harmonic motion of
the center of mass \cite{Kohn,Song05}. The system of a
noninteracting gas is invariant under the extended STT, which
predicts a generalized coherent (squeezed) state of breathing
(compressional) motion \cite{Song05,Sutherland,MG}. In experiments,
the system of the (generalized) {\em coherent state} can be prepared
by exciting the system in an eigenstate through shifting the trap
center and modulating the trap frequency. For the one-dimensional
boson gas of infinitely strong repulsive interaction of zero-range
(Tonks-Girardeau gas), the particle density is identical to that of
a noninteracting Fermi gas, and the invariance under the STT may be
applicable to some extent \cite{Song05,MG}. Recently invariance
under the STT has been used to analytically explain the "dynamical
fermionization"  of the momentum distribution of the expanding
Tonks-Girardeau gas \cite{TGexp,RM}.
%, based on stationary momentum
%distribution of the noninteracting Fermi gas \cite{MG}.

In this paper, exact and closed-form expressions for the particle
density, the kinetic energy density, the probability current
density, and the momentum distribution are presented for a state
{\em coherently excited} from the $(M+1)-$shell filled ground state
of a noninteracting Fermi gas in a $d$-dimensional isotropic
harmonic trap, and conservation laws for the relations of the
densities are given. The particle density and the momentum
distribution, which have been studied as a function of the
scattering length through a mean-field theory in
Ref.~\cite{Stringari}, are of great experimental interest
\cite{TGexp,RGGHJ}, and the expressions presented here are also
valid for the analysis of a sudden and total opening of the trap.
The profile of the momentum distribution turns out to be identical
in shape with that of the particle density, however, as a
manifestation of the Heisenberg {\em uncertainty principle}, the
dispersion of the distribution increases (decreases) when that of
the particle density is decreased (increased). The dispersion of the
distribution could be {\em arbitrarily} large or small, as the
amplitude of the breathing motion could be \cite{Song05}. Though
only the coherent states excited from a ground state will be
considered here, since the large or small dispersion basically comes
from the invariance under the STT and the uncertainty principle,
{\em general} coherent states could {\em also} have the large or
small dispersion. After a total opening of the trap, we show that
the momentum distribution of the free gas remains stationary without
changing from that of the gas at the moment of the opening. Since we
ignore the interactions, we consider only the spin-polarized case.

For a $d$-dimensional isotropic harmonic oscillator described by the
Hamiltonian
\begin{equation}
H(\mathbf{r},\mathbf{p},t)=\sum_{i=1}^d \left(\frac{{p_i}^2}{2 m}
+\frac{1}{2} m w^2(t){x_i}^2 -F_i(t)x_i\right)
%H_i(x_i,p_i,t),
\end{equation}
%with $\bold{r}=(x_1,x_2,\ldots, x_d)$, $\bold{p}=(p_1,p_2,\ldots,
%p_d)$, $\bold{n}=(n_1,n_2,\ldots, n_d)$, and
with $m$,  $w(t)$, and $F_i(t)$ denoting the positive mass, real
frequency, and an external force in the $i$-th direction,
respectively, through the extension of the one-dimension result
\cite{KL}, a complete set of the wave functions
$\psi_{\mathbf{n}}(\mathbf{r},t)$ satisfying the time-dependent
Schr\"{o}dinger equation is found, with $\mathbf{n}=(n_1,n_2,\ldots,
n_d)$, as
\begin{equation}
\psi_{\mathbf{n}}(\mathbf{r},t)=\prod_{i=1}^d
\psi_{n_i}(x_i,x_i^p,t),
\end{equation}
where $n_i$ is a non-negative integer, and
\begin{eqnarray}
&&\psi_{n_i}(x_i,x_i^p,t)\cr
 && = \frac{1}{\sqrt{2^{n_i} {n_i}!}\pi^{1/4} \sqrt{l}}
     \left[\frac{u-iv}{\eta(t)}\right]^{n_i+{1\over 2}}
     \exp\left[\frac{i}{\hbar}\left(\delta_i+m\dot{x}_i^p x_i\right) \right]
 \cr&&
  ~~\times \exp{\left[\frac{(x_i-x_i^p)^2}{2}\left(-\frac{1}{l^2}
               +i \frac{m\dot{l}}{\hbar l}\right)\right]}
  H_{n_i}( \frac{x_i -x_i^p}{l}),~~~~
\end{eqnarray}
with $u,~v$ being two linearly independent real homogeneous
solutions of the classical equation of motion
\begin{equation}
\ddot{x}_{cl}+w^2(t)x_{cl}=\frac{F_i(t)}{m},
\end{equation}
and $\eta(t)$, $l$  being defined as
\begin{equation}
\eta(t)=\sqrt{u^2 +v^2},~~~~~~~l=\eta\sqrt{\frac{\hbar}{\Omega}},
\end{equation}
respectively, while a positive constant $\Omega$ is defined as $
\Omega = m[ \dot{v}u - \dot{u}v]$, and the overdots denote
differentiations with respect to $t$. In Eq.~(3), $H_{n_i}$ is the
Hermite polynomial, $x_i^p$ denotes a solution of Eq.~(4), and
$\delta_i$ is defined through $
\dot{\delta}_i=\frac{m}{2}\left[w^2(t)x_i^p-\dot{x}_i^p\right].$

When $w(t)$ is a constant $w_c$ and $F_i(t)=0$, if we choose
$u=\cos(w_ct)$, $v=\sin(w_ct)$, and $x_i^p=0$, then
$l=\sqrt{\hbar/(mw_c)}$, and $\psi_{\mathbf{n}}(\mathbf{r},t)$ is
the wave function of an eigenstate. The $(M+1)-$shell filled
coherent state can be obtained from the $(M+1)-$shell filled ground
state, by applying an external force $F_i(t)$ and modulating $w(t)$.
If the coherent state is established once, the system  remains a
coherent state in general even after the external force is turned
off and the modulation is stopped \cite{Song05}. The one-particle
reduced density matrix of the $(M+1)-$shell filled coherent state is
given by
\begin{equation}
\rho(\mathbf{r}',\mathbf{r};t)=\sum_{{\{\mathbf{n}|\sum_i n_i \leq M}\}}
\psi_{\mathbf{n}}^*(\mathbf{r'},t)\psi_{\mathbf{n}}(\mathbf{r},t).
\label{eq:matrix}
\end{equation}

With an independent variable $b$, we define
\begin{equation}
C(\mathbf{r}',\mathbf{r};t;b)=\sum_{\{\mathbf{n}\}}
e^{-b\left(d/2+\sum_{i=1}^d n_i\right)}
\psi_{\mathbf{n}}^*(\mathbf{r}',t)\psi_{\mathbf{n}}(\mathbf{r},t).
\label{eq:Green}
\end{equation}
Making use of this fact
\begin{eqnarray}
&&\exp\left[-\frac{1}{2}\left(\tilde{x}^2+\tilde{y}^2\right)\right]
\sum_{n=0}\frac{e^{-b(1/2+n)}}{2^nn!}H_n(\tilde{x})H_n(\tilde{y})\cr
&&= \frac{\exp\left[
-\frac{(\tilde{x}+\tilde{y})^2}{4}\tanh\left(\frac{b}{2}\right)
-\frac{\left(\tilde{x}-\tilde{y}\right)^2}{4}\coth\left(\frac{b}{2}\right)
\right]}{\sqrt{2\sinh(b)}}
\end{eqnarray}
which comes from Mehler's formula \cite{Mehler},
we find that
\begin{eqnarray}
&&C(\mathbf{r'},\mathbf{r};t;b)\cr
&&=\frac{1}{(2\pi)^{d/2}l^d\sinh^{d/2}(b)}\cr
&&~\times\exp\left[i\frac{m}{\hbar}\left({\dot{\mathbf{r}}}_p\cdot\mathbf{s}
   +\frac{\dot{l}}{l}\mathbf{s}\cdot(\mathbf{q}-{\mathbf{r}}_p)\right)\right]\cr
&&~\times\exp\left[-\frac{(\mathbf{q}-\mathbf{r}_p)^2}{l^2}\tanh\left(\frac{b}{2}\right)
-\frac{{\mathbf{s}}^2}{4l^2}\coth\left(\frac{b}{2}\right)\right],~~~~~
\end{eqnarray}
where
\begin{equation}
\mathbf{q}=(\mathbf{r}+\mathbf{r'})/2,~~\mathbf{s}=\mathbf{r}-\mathbf{r'},
\end{equation}
with $\mathbf{r}_p=(x_1^p,x_2^p,\ldots, x_d^p)$. Just as in
Ref.~\cite{BZ}, the infinite summation in Eq.~(\ref{eq:Green}) can
be truncated, through the well-known facts in the (inverse) Laplace
transformation
\[
{\cal L}_b\left[\Theta(\beta-k)\right]=e^{-kb}/{b},~~~ {\cal
L}_\beta^{-1}\left[e^{-kb}/{b}\right]=\Theta(\beta-k),
\]
with a positive constant $k$ and the step function $\Theta(s)$
satisfying $\Theta(s)=1$ for $s>0$ and $\Theta(s)=0$ for $s<0$. The
density matrix may thus be formally written as
\begin{equation}
\rho(\mathbf{r}',\mathbf{r};t)={\cal L}_\lambda^{-1}
\left[\frac{1}{b}C(\mathbf{r}',\mathbf{r};t;b)\right],
\label{eq:densityformal}
\end{equation}
with $ \lambda=M+(d+1)/2$. By explicitly carrying out the inverse
Laplace transformation for the diagonal component of the density
matrix, we find the particle density
\begin{equation}
\rho(\mathbf{r},t)=\rho(\mathbf{r},\mathbf{r};t)=\frac{1}{l^d} {\cal
D}(z),
\end{equation}
where
\begin{equation}
z= (\mathbf{r}-\mathbf{r}_p)^2/{l^2}
\end{equation}
and the function ${\cal D}(z)$ is defined as \cite{BZ}
\begin{equation}
{\cal D}(z)=\frac{e^{-z}}{\pi^{d/2}}\sum_{n=0}^M
(-1)^nF_{M-n}^{(d)}L_n(z),
\end{equation}
with $ F_{2n}^{(d)}=(d/2+2n)\Gamma(d/2+n)/[n!(d/2)!],$ $
F_{2n+1}^{(d)}=2\Gamma(d/2+n+1)/[n!(d/2)!]$ and $L_n(z)$ being the
Laguerre polynomial.

The probability current density defined by
\begin{equation}
\bold{S}(\mathbf{r},t)=\frac{\hbar}{2im} {\sum}'
\left(\psi_{\mathbf{n}}^*(\mathbf{r},t)\nabla\psi_{\mathbf{n}}(\mathbf{r},t)
-\psi_{\mathbf{n}}(\mathbf{r},t)\nabla\psi_{\mathbf{n}}^*(\mathbf{r},t)\right)
\end{equation}
with ${\sum}'$ denoting the truncated summation of
Eq.~(\ref{eq:matrix}), can be found as
\begin{equation}
\mathbf{S}(\mathbf{r},t) = \frac{\hbar}{im}\nabla_\mathbf{s}
\rho(\mathbf{r}',\mathbf{r};t)|_{\mathbf{s}\rightarrow 0}=\mathbf{v}\rho(\mathbf{r},t),
\end{equation}
where
\begin{equation}
\mathbf{v}=\dot{\mathbf{r}}_p+(\dot{l}/l)(\mathbf{r}-\mathbf{r}_p).
\end{equation}
Up to the phase, the wave function $\psi_{\mathbf{n}}(\mathbf{r},t)$
can be obtained from that of an eigenstate by isotropically
rescaling the space according to the ratio $l$ and globally
displacing it by the amount $\mathbf{r}_p$.
%$\rho(\mathbf{r},t)$ can thus be read from the particle density of
%the ground state \cite{Song05,MG}.
Since, under the rescaling, $\dot{l}/l$ plays the same role as the
Hubble constant in the expanding universe, $\mathbf{v}$ is the
velocity of the position $\mathbf{r}$ under the transformations,
which explains the classical aspect of $\mathbf{S}(\mathbf{r},t)$.
% In this respect, the probability current density can be
% understood classically.

For the kinetic energy density, we explore the quantity
\begin{eqnarray}
&&\tau_\mathbf{s}(\mathbf{r},t)\cr &&=-\frac{\hbar^2}{8m}{\sum}'
[(\nabla^2\psi_{\mathbf{n}}^*(\mathbf{r},t))\psi_{\mathbf{n}}(\mathbf{r},t)
+\psi_{\mathbf{n}}^*(\mathbf{r},t)\nabla^2\psi_{\mathbf{n}}(\mathbf{r},t)
\cr &&~~~~~~~~~~~~~~~~~
-2(\nabla\psi_{\mathbf{n}}^*(\mathbf{r},t))\cdot\nabla\psi_{\mathbf{n}}(\mathbf{r},t)],
\end{eqnarray}
which can be evaluated as
\begin{eqnarray}
&&\tau_\mathbf{s}(\mathbf{r},t)=-\frac{\hbar^2}{2m}{\cal L}_\lambda^{-1}
\left.\left[\frac{1}{b}\nabla_\mathbf{s}^2C(\mathbf{r}',\mathbf{r};t;b)\right]
\right\vert_{\mathbf{s}\rightarrow 0}\nonumber\\
&&=\frac{m\mathbf{v}^2}{2}\rho(\mathbf{r},t)+\tau_Q(z,t),
\end{eqnarray}
where
\begin{equation}
\tau_Q(z,t)=\frac{d(2\pi)^{-d/2}\hbar^2}{4ml^{d+2}}
{\cal L}_\lambda^{-1}\left[\frac{\coth(\frac{b}{2})\exp[-z\tanh(\frac{b}{2})]}{b\sinh^{d/2}(b)}
\right].
\end{equation}
By carrying out the inverse Laplace transformation explicitly, we find that
\begin{equation}
\tau_Q(z,t)=\frac{\hbar^2de^{-z}}{4m\pi^{d/2}l^{d+2}}
\sum_{n=0}^M(-1)^nG_{M-n}^{(d)}L_n(z),
\end{equation}
with
$G_{2n}^{(d)}=(32n^2+d^2+16nd+2d)\Gamma(d/2+n)/[4(n!)\Gamma(d/2+2)],$
and $G_{2n+1}^{(d)}=2(4n+d+2)\Gamma(d/2+n+1)/[n!\Gamma(d/2+2)]$
\cite{BZ}. $\tau_Q(z,t)$ can be represented in terms of
$\rho(\mathbf{r},t)$ as
\begin{eqnarray}
&&\tau_Q(z,t)=\cr
&&\frac{d}{d+2}\frac{\hbar^2}{m}
\left[ \frac{1}{8}\mathbf{\nabla}^2\rho(\mathbf{r},t)
   +\frac{\left(M+\frac{d+1}{2}-\frac{z}{2}\right)}{l^2}\rho(\mathbf{r},t)\right],~~~~~
\end{eqnarray}
which, for the ground state, reduces to the known relation \cite{BZ,HMN}.
An expression of the kinetic energy density is then written as
\begin{eqnarray}
\tau(\mathbf{r},t)&=&\frac{\hbar^2}{2m}{\sum}'
[(\nabla\psi_{\mathbf{n}}^*(\mathbf{r},t))\cdot\nabla\psi_{\mathbf{n}}(\mathbf{r},t)]\cr
&=&\frac{m\mathbf{v}^2}{2}\rho(\mathbf{r},t)+2\frac{d+1}{d}\tau_Q(z,t)\cr
&&-\frac{\hbar^2}{ml^2}\left(M+\frac{d+1}{2}-\frac{z}{2}\right)\rho(\mathbf{r},t).
\end{eqnarray}

$\rho(\mathbf{r}',\mathbf{r};t)$ satisfies the following Schr\"{o}dinger equation
\begin{eqnarray}
&&i\hbar\frac{\partial }{\partial t}\rho(\mathbf{r}',\mathbf{r};t)\cr
&&=-\frac{\hbar^2}{2m}(\nabla^2-{\nabla'}^2) \rho(\mathbf{r}',\mathbf{r};t)\cr
   & &~~~+\left[\frac{mw^2(t)}{2}(\mathbf{r}^2-{\mathbf{r}'}^2)
                     -(\mathbf{r}-\mathbf{r}')\cdot\mathbf{F}(t)\right]
     \rho(\mathbf{r}',\mathbf{r};t).~~~~~
\label{eq:densitySch}
\end{eqnarray}
which can be used to find conservation laws \cite{conservation}. The
coincidence limit $\mathbf{r}\rightarrow\mathbf{r}'$ of
Eq.~(\ref{eq:densitySch}) gives
\begin{equation}
\frac{\partial}{\partial t}\rho(\mathbf{r},t)+\nabla\cdot[\rho(\mathbf{r},t)\mathbf{v}]=0.
\end{equation}
The  coincidence limit after applying $\nabla_{\mathbf{s}}$ on
Eq.~(\ref{eq:densitySch}) yields
\begin{eqnarray}
&&\frac{\partial}{\partial t}\left[\rho(\mathbf{r},t)\mathbf{v}_i\right]
+\frac{\partial}{\partial x_j}\left[ \rho(\mathbf{r},t)\mathbf{v}_i\mathbf{v}_j\right]
+\frac{4x_i}{mdl^2}\frac{\partial}{\partial z}\tau_Q(z,t)\cr
&&=-\rho(\mathbf{r},t)[w^2(t)x_i-\frac{F_i(t)}{m}].
\label{eq:knowncon}
\end{eqnarray}
Similarly, by applying $\nabla_{\mathbf{s}}^2$ on
Eq.~(\ref{eq:densitySch}) with the help of
Eq.~(\ref{eq:densityformal}), we find that
\begin{equation}
\frac{\partial}{\partial t}\tau_Q(z,t)
+\frac{\partial}{\partial x_i}\left[\tau_Q(z,t)\mathbf{v}_i\right]=-2\frac{\dot{l}}{l}\tau_Q(z,t).
\end{equation}
For the ground state of $w(t)=w_c$ and $F_i(t)=0$, Eq.~(\ref{eq:knowncon}) reduces to the known relation
\cite{ZBSB}.

The momentum distribution defined by
\begin{equation}
n(\mathbf{p},t)=\frac{1}{(2\pi)^d}\int d\mathbf{r}d\mathbf{r}'
e^{-i\mathbf{p}\cdot(\mathbf{r}-\mathbf{r}')/\hbar}\rho(\mathbf{r}',\mathbf{r};t)
\end{equation}
can be calculated as
\begin{eqnarray}
n(\mathbf{p},t)&=&\frac{\alpha^{-d}(t)}{(2\pi)^{d/2}}{\cal
L}_\lambda^{-1}
\left[\frac{e^{-\left[(\mathbf{p}-m\dot{\mathbf{r}}_p)^2
\tanh(\frac{b}{2})/\alpha^2(t)\right]}}{b\sinh^{d/2}(b)}\right]\cr
&=&\alpha^{-d}(t){\cal
D}\left((\mathbf{p}-m\dot{\mathbf{r}}_p)^2/\alpha^2(t)\right),
\end{eqnarray}
where
\begin{equation}
\alpha(t)=\sqrt{\hbar^2/l^2+m^2\dot{l}^2}. \label{eq:alpha}
\end{equation}
The momentum distribution in the momentum space thus has the
identical shape of the particle density in coordinate space upon
displacement and rescaling, as has been estimated for the
one-dimensional free expanding gas \cite{MG}. Just as in the
particle density (see, e.g., \cite{VMT,BZ,Mueller}), with the
particle number $N$, the momentum distribution goes over to the
%local density (Thomas-Fermi)
approximate form
\begin{equation}
n_{TF}=\frac{\left[2(d!N)^{1/d}-\alpha^{-2}(t)
   (\mathbf{p}-m\dot{\mathbf{r}}_p)^2\right]^{d/2}}{2^d\pi^{d/2}\Gamma(1+d/2)\alpha^d(t)},
\label{eq:TF}
\end{equation} for $2(d!N)^{1/d}>
(\mathbf{p}-m\dot{\mathbf{r}}_p)^2/\alpha^2(t)$, in the large $N$
limit, and the exact profile has corrugation-like corrections from
the approximation \cite{Mueller,BZ}. For a time-constant trap, if we
choose $u=\cos(w_ct)$, $v=\sin(w_ct)$, and $x_i^p=0$, $n_{TF}$
exactly agrees with the distribution found  for the deep BCS limit
of the atomic Fermi gas through the mean-field theory
\cite{Stringari} (see also Ref.~\cite{RGGHJ}).

For a general $w(t)$, $l$ satisfies
$m\ddot{l}+mw^2(t)l-\hbar^2/(ml^3)=0$, and the behavior of $l$ has
been analyzed for time-periodic $w(t)$ \cite{KL} to find that $l$
could be more and more amplified in general and become unstable as
time passes. For a time-constant trap, as in the case after the
modulation of the frequency is stopped, $l$ satisfies $
m^2\dot{l}^2+\hbar^2/l^2=-m^2w_c^2l^2+2m\varepsilon=\alpha^2(t)$
with a constant $\varepsilon$, which shows, {\em as a manifestation
of the uncertainty principle} \cite{uncertainty}, $\alpha(t)$,
characterizing the dispersion of the momentum distribution,
increases (decreases) when $l$ or the dispersion of the particle
density is decreased (increased). For the time-constant trap, up to
a global time-displacement, $l$ is written without losing generality
as $l=\sqrt{\hbar/(mw_c)}\sqrt{A^2\cos^2 w_ct+\sin^2 w_ct/A^2}$,
with a positive constant $A$ \cite{Song05}. In this case, $\alpha(t)
=\sqrt{m\hbar w_c[A^2\sin^2 w_ct+\cos^2 w_ct/A^2]}$, which shows,
the dispersion of the momentum distribution could be {\em
arbitrarily large or small} in the limit of $A\rightarrow 0$ or
$A\rightarrow \infty$.

The formalisms developed are also valid for $w(t)=0$ in
$d$-dimension, with $u=a_1$, $v=a_2(t-t_0)+a_3$ of $a_1,a_2,a_3,t_0$
being constants. For a sudden and total opening of the trap at
$t=t_0$, the continuity of the wave functions requires the
continuities of $x_i^p,~\dot{x}_i^p,~\eta,~\dot{\eta}$, which gives
%\begin{equation}
%a_1^2+a_3^2=\eta_{-}^2,~~~a_2a_3=\eta_{-}\dot\eta_{-}, ~~~a_1a_2=\Omega/m,
%\label{eq:continuity}
%\end{equation}  (\ref{eq:continuity})
%is solved to give
$ a_2=\sqrt{\Omega^2/(m\eta_{-})^2+\dot{\eta}_{-}^2},~~
a_1=\frac{\Omega}{ma_2},~~a_3=\frac{\eta_{-}\dot\eta_{-}}{a_2}, $
where the subscript $-$ denotes that the quantity is evaluated in
the limit of $t-t_0\rightarrow -0$. If $\dot\eta_{-}<0$, the
dispersion of the particle density, which is proportional to $l$,
will decrease after the opening of the trap until $t-t_0=|a_3|/a_2$,
and then it will increase, while the cloud  of the gas will
continuously expand for $\dot\eta_{-}\geq0$. $\alpha(t)$ of
Eq.~(\ref{eq:alpha}) with
$l=\sqrt{\hbar[a_1^2+(a_2(t-t_0)+a_3)^2]/(ma_1a_2)}$ is a constant
$\alpha_-$, that is, for $t>t_0$, $\alpha(t)=\sqrt{m\hbar
a_2/a_1}=\sqrt{\hbar^2/l_-^2+m^2\dot{l}_-^2 }=\alpha_-.$ After the
opening, {\em the momentum distribution of the gas is thus
stationary}, as has been predicted for the one-dimensional expanding
gas. If the opening is for the ground state of the trap of a
constant frequency $w_c$, $l_-=\sqrt{\hbar/(mw_c)}$, and
$\dot\l_-=0$, and thus $\alpha(t)=\sqrt{\hbar mw_c}$ \cite{MG}.

In summary, the expressions of the particle, kinetic energy,
probability current densities, and the momentum distribution have
been derived for a coherent state of a noninteracting Fermi gas in a
$d$-dimensional isotropic harmonic trap, and, for the relations of
the densities, conservation laws have been given.
%The probability current density clearly shows classical
%aspects of the motion of the quantum fluid, and conservation laws
%for the relation of the densities are given.
It has been shown that the profile of the momentum distribution is
identical in shape with that of the particle density, while the
dispersion of the distribution increases (decreases) when that of
the particle density is decreased (increased) in a time-constant
trap. The dispersion of the distribution could be arbitrarily large
or small, which  basically comes from the invariance under the STT
and the uncertainty principle. Since the invariance exists only for
restricted cases \cite{Song05,Sutherland}, an appearance of the
distribution of the large or small dispersion in a system of Fermi
atoms may indicate that the system is in the noninteracting regime.
After a total opening of the trap, the momentum distribution does
not change from that of the gas at the moment of the opening, while,
long after the opening, the dispersion of the particle density
increases according to $\alpha_-t/m$ with the constant $\alpha_-$
which characterizes the dispersion of the distribution. As the
uncertainties are {\em controlled by the classical solutions} in the
coherent states, an experimental realization of the system described
here may be important to provide an {\em observable} manifestation
of the uncertainty principle.

This work was supported by the Korea Research Foundation Grant
funded by the Korean Government (MOEHRD) (KRF-2005-015-C00115).


\begin{thebibliography}{150}
\bibitem{uncertainty} The momentum distribution is a summation of
the momentum probability densities of the coherent states,
% in the momentum space of the
%coherent states $(1/(2\pi))^{-d/2}\int
%d\mathbf{r}e^{-i\mathbf{p}\cdot\mathbf{r}/\hbar}\psi_{\mathbf{n}}(\mathbf{r},t)$,
%for $\sum_i n_i \leq M$,
while the momentum uncertainty of a coherent state is inversely
proportional to the position uncertainty of the state.
\end{thebibliography}
\end{document}